\documentstyle[amstex,preprint,aps]{revtex}

\begin{document}
\title{Quantum Computing Using an Open System and Projected Subspace}
\author{Bi Qiao$^{ab}$, Harry. E. Ruda$^{a}$ and X. H. Zhen$^{c}$}
\address{$^{a}$Energenius Centre for Advanced Nanotechnology, University of\\
Toronto, Canada M5S 3E4.\\
$^{b}$Wuhan Institute of Physics and Mathmatics, Chinese Academy of Science,%
\\
Wuhan 430071, P. R. China\\
$^{c}$Complexity Science Center, Yangzhou University, Yangzhou 225002, P. R.%
\\
China}
\maketitle

\begin{abstract}
Using the subdynamical kinetic equation for an open quantum system, a
formulation is presented for performing decoherence-free (DF) quantum
computing in Rigged Liouville Space (RLS). Three types of interactions were
considered, and in each case, stationary and evolutionary states were
evaluated for DF behavior in both the total space and the projected
subspace. Projected subspaces were found using the subdynamics kinetic
equation. It was shown that although the total space may be decoherent, the
subspace can be DF. In the projected subspace, the evolution of the density
operator may be time asymmetric. Hence, a formulation for performing quantum
computing in RLS or rigged Hilbert space (RHS) was proposed, and a quantum
Controlled-Not Logical gate with corresponding operations in RLS (RHS) was
constructed. A generalized quantum Turing machine in RHS was also discussed.



{\Large \ }

{\LARGE \ }
\end{abstract}

\section{Introduction}

Although an ideal quantum computing system is an isolated quantum system
whose evolution is described by a unitary operator with time reversal
symmetry, in practice, it is often not practical to isolate a real quantum
computing system from its environment. The interaction between a real system
and its environment introduces decoherence which destroys the superpositions
of qubits that enable quantum logical operations to be validated$^{\left[ 1,2%
\right] }$. Several authors have formulated theories for decoherence-free
(DF) subspaces in which quantum computing can be performed. Theory is
developed starting from a master equation (such as the Lindblad equation)
for an open system within the powerful semigroup approach. This shows that
such DF subspaces do indeed exist, allowing logical qubits to be encoded and
not to decohere$^{\left[ 3,4\right] }$ under the Born-Markov approximation
and with restrictions on the type of decoherence (i.e., symmetric or
collective decoherence). This motivated us to identify appropriately
designed open quantum computing systems useful for canceling the effects of
intrinsic decoherence. Such systems would thus behave as ideal quantum
computing systems in the appropriate subspaces without introducing
approximations or restrictions on the type of decoherence. In such open
systems, self-adjoint operators and unitary evolution groups are not
intrinsically necessary to govern or operate quantum computation. Quantum
computation can then be performed in a more general functional space which
permits irreversibility, such as Rigged Hilbert Space (RHS) or Rigged
Liouville Space (RLS), rather than just Hilbert or Liouville Space. These
are kinds of triplet space structure, such as the dense subspace and its
topological dual space rigged to Hilbert space, $\Phi \subset {\cal H}%
\subset \Phi ^{\times }$, (RHS) or dense space and its topological dual
space rigged to Liouville space, $\Phi \otimes \Phi \subset {\cal H}\otimes 
{\cal H}\subset $ $^{\times }\Phi \otimes \Phi ^{\times }$, (RLS)$^{\left[
5,6\right] }$. These generalized functional spaces were introduced in
physics in order to provide insight into singular entities like Dirac's
delta functions and formulations of quantum mechanics$^{\left[ 7-8\right] }$%
. Since then, they have been intensively developed for describing
irreversible process, non-integrable systems and chaotic dynamical systems.
The latter includes RHS of the Hardy class used to formulate the generalized
eigenvectors and complex spectral decomposition of the Liouvillian for the
Friedrichs model$^{\left[ 9-11\right] }$ and for the general spectral
decomposition of chaotic maps$^{\left[ 6\right] }$. Such an approach can
also be used to formulate open quantum systems in a suitable functional
space, revealing non-unitary evolution semigroup irreversibility. It may
also be suitable for describing quantum logical operations in open quantum
systems subject to decoherence.

Based on the above concepts, we propose a type of projected subspace for
open quantum computing systems, to perform DF quantum logical operations for
three different types of interactions, and investigate quantum logical
operations in such subspaces based on the kinetic equation introduced from
subdynamics$^{\left[ 6,12\right] }$. Compared with recent publications on DF
subspaces, our approach is general and not restricted to the Born-Markov
approximation. In section II, a subdynamic formulation and kinetic equation
for an open system is derived using a simpler approach than previously
published references$^{\left[ 6,12\right] }$. In sections III, IV and V,
three types of coherence / DF conditions for the projected subspaces or
total spaces are analyzed under respective interactions. The role of
eigenvalues and eigenvectors of the Hamiltonian in the decoherent / DF
processes of the systems are discussed. In sections VI and VII, a
formulation for quantum logical operations in the projected space is
investigated, and a formulation of quantum computation in RLS (RHS) and a
generalized Turing machine are proposed.

\section{Subdynamical Kinetic Equation}

In general, consider an open quantum computing system $S$, which may consist
of real spins, pseudospins, or number states of a quantized field (such as
photons), coupled with an infinite or finite environment $R$, and where the
Hamiltonian operator for the total system is given by 
\[
H\left( t\right) =H_{0}\left( t\right) +\lambda H_{1}\left( t\right) , 
\]%
and $H_{0}\left( t\right) \equiv H_{S}\left( t\right) +H_{R},H_{S}\left(
t\right) $, $H_{R}$ and $H_{I}\left( t\right) $ correspond to the
Hamiltonian of $S$, $R$ and the interaction between $S$ and $R$. $\lambda $
is a coupling constant for this interaction. The corresponding Liouville
equation for the total density operator $\rho \left( t\right) $ is given by 
\[
i\frac{d}{dt}\rho \left( t\right) =\left[ H_{0}\left( t\right) +\lambda
H_{1}\left( t\right) ,\rho \left( t\right) \right] =\left( L_{0}\left(
t\right) +\lambda L_{1}\left( t\right) \right) \rho \left( t\right) . 
\]

Choosing the time-independent eigenprojectors of $L_{0}\left( t\right) $ as $%
P_{\nu }$ and $Q_{\nu }$ with $Q_{\nu }+P_{\nu }=1,$ respectively, then the
eigenprojectors $\Pi _{\nu }\left( t\right) $ for the total Liouvillian $%
L\left( t\right) $ can be written in terms of the Heisenberg equation as $i%
\frac{\partial }{\partial t}\Pi _{\nu }\left( t\right) =\left[ L\left(
t\right) ,\Pi _{\nu }\left( t\right) \right] $. $\Pi _{\nu }\left( t\right) $
satisfy the usual properties of projection operators, such as $\sum_{\nu
}\Pi _{\nu }\left( t\right) =1$, $\Pi _{\nu }^{2}\left( t\right) =1$, $\Pi
_{\nu }\left( t\right) \Pi _{\mu }\left( t\right) =\delta _{\nu \mu }$, and
is analytic with respect to $\lambda $: lim$_{\lambda \longrightarrow 0}\Pi
_{\nu }\left( t\right) \longrightarrow P_{\nu }$. From the definition of the
eigenprojectors $\Pi _{\nu }\left( t\right) $ we have $Q_{\nu }L\left(
t\right) \left( Q_{\nu }+P_{\nu }\right) \Pi _{\nu }\left( t\right) =Z_{\nu
\alpha }Q_{\nu }\Pi _{\nu }\left( t\right) $ ($Z_{\nu \alpha }$ is a
corresponding eigenvalue), giving%
\begin{equation}
Q_{\nu }\Pi _{\nu }\left( t\right) =C_{\nu }\left( t\right) \Pi _{\nu
}\left( t\right) ,  \label{eqn1}
\end{equation}%
where the creation operator is defined by $C_{\nu }\left( t\right) \equiv 
\mathrel{\mathop{\sum }\limits_{\alpha }}%
C_{\nu }\left( Z_{\nu \alpha },t\right) P_{\nu }\Pi _{\nu }P_{\nu }\ $with $%
C_{\nu }\left( z,t\right) \equiv \frac{1}{z-Q_{\nu }L\left( t\right) Q_{\nu }%
}Q_{\nu }L\left( t\right) P_{\nu }$, for $z\in $ complex plane ${\cal C}$.
In the same way, the destruction operator $D_{\nu }\left( t\right) $ is
defined by 
\begin{equation}
\Pi _{\nu }\left( t\right) Q_{\nu }=\Pi _{\nu }\left( t\right) D_{\nu
}\left( t\right) .  \label{eqn2}
\end{equation}%
Notice that, since $\Pi _{\nu }$ is the projector, therefore $Q_{\nu }\neq
C_{\nu }\left( t\right) \neq D_{\nu }\left( t\right) $ in Eqs.(\ref{eqn1})
and (\ref{eqn2}). Here $C_{\nu }\left( z,t\right) $ is the collision operator%
$^{\left[ 13,14\right] }$ used in non-equilibrium statistical mechanics. The
operator $C_{\nu }\left( t\right) $ creates the $Q_{\nu }$-part of $\Pi
_{\nu }\left( t\right) $ from $P_{\nu }$ and the operator $D_{\nu }\left(
t\right) $ provides the opposite operation of the creation operator since $%
C_{\nu }\left( t\right) =Q_{\nu }C_{\nu }\left( t\right) P_{\nu }$ and $%
D_{\nu }\left( t\right) =P_{\nu }D_{\nu }\left( t\right) Q_{\nu }$. This
allows the construction of kinetic equations for each $P_{\nu }$-component
of $\Pi _{\nu }\left( t\right) $ by 
\begin{eqnarray}
i\frac{\partial }{\partial t}\left( P_{\nu }\Pi _{\nu }\left( t\right) \rho
\left( t\right) \right) &=&iP_{\nu }\left[ \frac{\partial }{\partial t}\Pi
_{\nu }\left( t\right) \right] \rho \left( t\right) +iP_{\nu }\Pi _{\nu
}\left( t\right) \frac{\partial }{\partial t}\rho \left( t\right)
\label{eqn3} \\
&=&P_{\nu }\left[ L\left( t\right) ,\Pi _{\nu }\left( t\right) \right] \rho
\left( t\right) +P_{\nu }\Pi _{\nu }\left( t\right) L\left( t\right) \rho
\left( t\right)  \nonumber \\
&=&P_{\nu }L\left( t\right) \left( P_{\nu }+Q_{\nu }\left( t\right) \right)
\Pi _{\nu }\left( t\right) \rho \left( t\right)  \nonumber \\
&=&P_{\nu }L\left( t\right) \left( P_{\nu }+C_{\nu }\left( t\right) \right)
P_{\nu }\Pi _{\nu }\left( t\right) \rho \left( t\right)  \nonumber \\
&=&\Theta _{\nu }\left( t\right) \left( P_{\nu }\Pi _{\nu }\left( t\right)
\rho \left( t\right) \right) ,  \nonumber
\end{eqnarray}%
where $\Theta _{\nu }\left( t\right) \equiv P_{\nu }L\left( t\right) P_{\nu
}+P_{\nu }L\left( t\right) C_{\nu }\left( t\right) P_{\nu }$. A kinetic
equation for all of $P_{\nu }$ is given%
\begin{eqnarray}
i\frac{\partial }{\partial t}\sum_{\nu }\left( P_{\nu }\Pi _{\nu }\left(
t\right) \rho \left( t\right) \right) &=&\sum_{\nu }\Theta _{\nu }\left(
t\right) \sum_{\mu }\left( P_{\mu }\Pi _{\mu }\left( t\right) \rho \left(
t\right) \right)  \label{eqn3aa} \\
&=&\Theta \left( t\right) \rho ^{proj}\left( t\right) ,  \nonumber
\end{eqnarray}%
where we define the total intermediate operator as $\Theta \left( t\right)
\equiv \sum_{\nu }\Theta _{\nu }\left( t\right) $ and the total projector as 
$\rho ^{proj}\left( t\right) \equiv \sum_{\nu }\rho _{\nu }^{proj}\left(
t\right) \equiv \sum_{\nu }P_{\nu }\Pi _{\nu }\left( t\right) \rho \left(
t\right) $. We refer to Eq.(\ref{eqn3aa}) as the subdynamical kinetic
equation, and the operator $\Theta \left( t\right) $ as the generator of the
subdynamical kinetic equation. The second-order approximation for each $%
\Theta _{\nu }\left( t\right) $ corresponds to the Bolzmann, Pauli, and
Fokker-Planck equations of kinetic theory and Brownian motion$^{\left[ 6,12%
\right] }$. The creation and destruction operators can be obtained from the
basic operator equations within the subdynamics formulation$^{\left[ 6,12%
\right] }$, 
\begin{eqnarray}
i\frac{\partial }{\partial t}C_{\nu }\left( t\right) +\left[ L_{0},C_{\nu
}\left( t\right) \right] &=&\lambda \left( C_{\nu }-Q_{\nu }\right)
L_{1}\left( P_{\nu }+C_{\nu }\right) ,  \label{eqn4} \\
i\frac{\partial }{\partial t}D_{\nu }\left( t\right) +\left[ L_{0},D_{\nu
}\left( t\right) \right] &=&\lambda \left( P_{\nu }+D_{\nu }\right)
L_{1}\left( Q_{\nu }-D_{\nu }\right) .  \label{eqn5}
\end{eqnarray}%
The construction of a kinetic equation for the whole system can been
accomplished by solving the equations for the $C_{\nu }\left( t\right) $, or 
$D_{\nu }\left( t\right) $, with retarded or advanced integrations; for
example, 
\begin{equation}
C_{\nu }\left( t\right) =-i\int_{t}^{\mp \infty }d\tau \widehat{T}%
e^{-i\int_{\tau }^{t}L_{0}\left( t^{\prime }\right) dt^{\prime }}\lambda
\left( C_{\nu }\left( \tau \right) -Q_{\nu }\right) L_{1}\left( \tau \right)
\left( P_{\nu }+C_{\nu }\left( \tau \right) \right) \widehat{T}%
e^{i\int_{\tau }^{t}L_{0}\left( t^{\prime }\right) dt^{\prime }},
\label{eqn6}
\end{equation}%
where $\widehat{T}$ is the Dyson time-ordered operator. A similar approach
can be used for the destruction operator. Neglecting higher terms $0\left(
C_{\nu }^{2}\left( t\right) \right) $, $0\left( D_{\nu }^{2}\left( t\right)
\right) $, Eqs.(\ref{eqn4}) and (\ref{eqn5}) become 
\begin{eqnarray}
i\frac{\partial }{\partial t}C_{\nu }\left( t\right) +\left[ L_{0},C_{\nu
}\left( t\right) \right] &=&-\lambda Q_{\nu }L_{1}\left( P_{\nu }+C_{\nu
}\right) ,  \label{eqn7} \\
i\frac{\partial }{\partial t}D_{\nu }\left( t\right) +\left[ L_{0},D_{\nu
}\left( t\right) \right] &=&-\lambda \left( P_{\nu }+D_{\nu }\right)
L_{1}Q_{\nu }.  \label{eqn8}
\end{eqnarray}%
The exact solutions for Eqs.(\ref{eqn7}) and (\ref{eqn8}) are: 
\begin{equation}
C_{\nu }\left( t\right) =-i\lambda e^{-i\int_{\pm \infty }^{t}Q_{\nu
}L\left( \tau ^{\prime }\right) Q_{\nu }d\tau ^{\prime }}\left[ \int_{\pm
\infty }^{t}e^{i\int_{\pm \infty }^{\tau }Q_{\nu }L\left( \tau ^{\prime
}\right) Q_{\nu }d\tau ^{\prime }}Q_{\nu }L\left( \tau \right) P_{\nu
}e^{-iP_{\nu }L_{0}P_{\nu }\tau }d\tau \right] e^{iP_{\nu }L_{0}P_{\nu }t},
\label{eqn9}
\end{equation}%
and 
\begin{equation}
D_{\nu }\left( t\right) =-i\lambda e^{iP_{\nu }L_{0}P_{\nu }t}\left[
\int_{\pm \infty }^{t}e^{-iP_{\nu }L_{0}P_{\nu }\tau }Q_{\nu }L\left( \tau
\right) P_{\nu }e^{i\int_{\pm \infty }^{\tau }Q_{\nu }L\left( \tau ^{\prime
}\right) Q_{\nu }d\tau ^{\prime }}d\tau \right] e^{-i\int_{\pm \infty
}^{t}Q_{\nu }L\left( \tau ^{\prime }\right) Q_{\nu }d\tau ^{\prime }}.
\label{eqn10}
\end{equation}

The general evolution formulae for the density operator can be deduced from
Eq.(\ref{eqn3}). Indeed, since 
\begin{eqnarray}
&&P_{\nu }\Pi _{\nu }\left( t\right) \rho \left( t\right) =\left( P_{\nu
}+D_{\nu }\left( t\right) C_{\nu }\left( t\right) \right) ^{-1}\left( P_{\nu
}+D_{\nu }\left( t\right) \right) \rho \left( t\right)  \label{eqn11} \\
&=&\left( P_{\nu }+D_{\nu }\left( t\right) C_{\nu }\left( t\right) \right)
^{-1}\left( P_{\nu }+D_{\nu }\left( t\right) \right) 
\mathrel{\mathop{\sum }\limits_{\nu ^{\prime }}}%
\left( P_{\nu ^{\prime }}+C_{\nu ^{\prime }}\left( t\right) \right) \rho
_{\nu ^{\prime }}\left( t\right)  \nonumber \\
&=&\delta _{\nu \nu ^{\prime }}\left( P_{\nu }+D_{\nu }\left( t\right)
C_{\nu }\left( t\right) \right) ^{-1}\left( P_{\nu ^{\prime }}+D_{\nu
^{\prime }}\left( t\right) C_{\nu ^{\prime }}\left( t\right) \right) \rho
_{\nu ^{\prime }}^{proj}\left( t\right)  \nonumber \\
&=&\rho _{\nu }^{proj}\left( t\right) ,  \nonumber
\end{eqnarray}%
and 
\begin{eqnarray}
\rho \left( t\right) &\equiv &%
\mathrel{\mathop{\sum }\limits_{\nu }}%
\left( P_{\nu }+C_{\nu }\left( t\right) \right) \rho _{\nu }^{proj}\left(
t\right)  \label{eqn12} \\
&=&\Omega \rho _{\nu }^{proj}\left( t\right) ,  \nonumber
\end{eqnarray}%
where $\Omega $ is the similarity operator defined as 
\begin{equation}
\Omega \equiv 
\mathrel{\mathop{\sum }\limits_{\nu }}%
\left( P_{\nu }+C_{\nu }\left( t\right) \right) .  \label{eqn13}
\end{equation}%
Eq.(\ref{eqn3}) can be understood as\qquad 
\begin{eqnarray}
i\frac{\partial }{\partial t}\rho _{\nu }^{proj}\left( t\right) &=&i\frac{%
\partial }{\partial t}P_{\nu }\Pi _{\nu }\left( t\right) \rho \left( t\right)
\nonumber \\
&=&\Theta _{\nu }\left( t\right) P_{\nu }\Pi _{\nu }\left( t\right) \rho
\left( t\right)  \label{eqn14} \\
&=&\Theta _{\nu }\left( t\right) \rho _{\nu }^{proj}\left( t\right) 
\nonumber \\
&=&\left( P_{\nu }LP_{\nu }+P_{\nu }L_{\nu }C_{\nu }\left( t\right) P_{\nu
}\right) \rho _{\nu }^{proj}\left( t\right) ,  \nonumber
\end{eqnarray}%
which gives the formal solution for $\rho _{\nu }^{proj}\left( t\right) $,

\begin{eqnarray}
\rho _{\nu }^{proj}\left( t\right) &=&\widehat{T}e^{-i\int_{t_{0}}^{t}dt^{%
\prime }\Theta _{n}\left( t^{\prime }\right) }\rho _{\nu }^{proj}\left(
t_{0}\right)  \label{eqn15} \\
&=&\widehat{T}e^{-i\int_{t_{0}}^{t}dt^{\prime }\Theta _{n}\left( t^{\prime
}\right) }\left( P_{\nu }+D_{\nu }\left( t_{0}\right) C_{\nu }\left(
t_{0}\right) \right) ^{-1}\left( P_{\nu }+D_{\nu }\left( t_{0}\right)
\right) \rho ^{proj}\left( t_{0}\right) .  \nonumber
\end{eqnarray}%
Eq.(\ref{eqn15}), for each $P_{\nu }$-component of $\Pi _{\nu }\left(
t\right) $ is related to the dynamics of $\rho _{\nu }^{proj}\left( t\right) 
$ and includes the master equation as the second-order approximation of $%
\Theta _{\nu }\left( t\right) $ with respect to weak coupling$^{\left[ 12%
\right] }$. The Liouville equation for the reduced projected density of the
quantum computing subsystem $S$ can be obtained by taking the part trace $%
Tr_{R}$ of Eq.(\ref{eqn15}). It can be shown that the Liouvillian $L\left(
t\right) $ and the total intermediate operator $\Theta \left( t\right) $
hold a similarity relation (i.e., in the time-dependent case, $L\left(
t\right) =\left( i\frac{\partial \Omega \left( t\right) }{\partial t}+\Omega
\left( t\right) \Theta \left( t\right) \right) \Omega ^{-1}\left( t\right) $%
). In the time independent case, $L=\Omega \Theta \Omega ^{-1}$, which can
replace the Hamiltonian and relevant projections$^{\left[ 15\right] }$.
Therefore, the eigenvectors of $\Theta $ in the time-independent case, can
be transformed to $L$ with the same structure of eigenvalues for $\Theta $
and $L$. Thus Eq.(\ref{eqn14}) or (\ref{eqn3aa}) is the general kinetic
equation with significant physical meaning, and can exactly describe the
dynamics of the open quantum computing system $S$ without invoking any
approximations. Hence, it provides the starting point for the discussions
below.

\section{Decoherence and DF Interactions in Total Space for a Diagonal
Interaction}

Using Eq.(\ref{eqn14}) or (\ref{eqn3aa}), it is aparent that the second term
on the right tends to zero corresponding to a complete DF subspace for the
bipartite system $S+R$. This suggests that one should try to identify
bipartite systems which, under ``diagonal interaction'' with the
environment, still permit the second term to be zero and remain coherent.
Here the term ``diagonal interaction'' is taken to mean that the
eigenfunctions of the interaction part of the Hamiltonian are still the
original product states of the unperturbed part of the Hamiltonian$^{\left[
16\right] }$. Indeed, if the Hamiltonian of an open system has a diagonal
interaction form, then the interaction term of Eq.(\ref{eqn14}) or (\ref%
{eqn3aa}), should equal zero and a DF condition exist for the stationary
states of the system.

To illustrate this point, consider a two-level atom interacting with a
single cavity mode. Although this model does not describe decoherence from a
Bosonic bath (since it only deals with one bosonic degree of freedom) it is
sufficient to show the underlying principles. In the case of an off-resonant
interaction, the effective Hamiltonian can be expressed as$^{\left[ 16,17%
\right] }$%
\begin{equation}
H=\omega _{0}\sigma ^{z}+\omega a^{+}a+ga^{+}a\left| 2\right\rangle
\left\langle 2\right| ,  \label{eqn16}
\end{equation}%
where $\sigma ^{z}\equiv \left| 2\right\rangle \left\langle 2\right| -\left|
1\right\rangle \left\langle 1\right| $, $\omega _{0}$ and $\omega $ are
angular frequencies, $g$ is the coupling number, and $a^{+}$, $a$ are
creation and annihilation operators for the cavity mode, respectively.

Without exercising the diagonal interaction, $ga^{+}a\left| 2\right\rangle
\left\langle 2\right| $, the spectral decomposition of the total Hamiltonian
for this system can be expressed as 
\begin{equation}
H_{0}=\sum_{n}\sum_{j=1}^{2}\left( \left( -1\right) ^{j}\omega _{0}+n\omega
\right) \left| j\left( 1\right) \otimes n\left( 2\right) \right\rangle
\left\langle n\left( 2\right) \otimes j\left( 1\right) \right|  \label{eqn17}
\end{equation}%
with 
\begin{equation}
\left| \psi \right\rangle =\sum_{n}\sum_{j=1}^{2}\left\langle n\left(
2\right) \otimes j\left( 1\right) \right| \left. \psi \right\rangle \left|
j\left( 1\right) \otimes n\left( 2\right) \right\rangle ,\text{ for any }%
\left| \psi \right\rangle \text{ of this system.}  \label{eqn18}
\end{equation}%
Exercising the diagonal interaction, $ga^{+}a\left| 2\right\rangle
\left\langle 2\right| $, the spectral decomposition of the total Hamiltonian
is 
\begin{equation}
H=\sum_{n}\sum_{j=1}^{2}\left( \left( -1\right) ^{j}\omega _{0}+n\omega +%
\frac{1+\left( -1\right) ^{j}}{2}ng\right) \left| j\left( 1\right) \otimes
n\left( 2\right) \right\rangle \left\langle n\left( 2\right) \otimes j\left(
1\right) \right| .  \label{eqn19}
\end{equation}%
It is therefore apparent that the eigenvectors of this Hamiltonian $H$ are
still the original eigenvectors of $H_{0}$ although the eigenvalues change
to $\left( -1\right) ^{j}\omega _{0}+n\omega +\frac{1+\left( -1\right) ^{j}}{%
2}ng$ from $\left( -1\right) ^{j}\omega _{0}+n\omega $. $H$ is diagonal with
respect to the basis $\left\{ \left| j\left( 1\right) \right\rangle \otimes
\left| n\left( 2\right) \right\rangle ,\left\langle n\left( 2\right) \right|
\otimes \left\langle j\left( 1\right) \right| \right\} $, and therefore the
off-diagonal matrix elements of the corresponding Liouvillian are equal to
zero, 
\begin{equation}
P_{\nu }L_{1}\left( t\right) Q_{\nu }=0,\text{ for }\nu =\left( j,j^{\prime
};n,n^{\prime }\right) ,  \label{eqn20}
\end{equation}%
Eq.(\ref{eqn20}) yields 
\begin{equation}
P_{\nu }L_{1}\left( t\right) Q_{\nu }C_{\nu }\left( t\right) P_{\nu }=0,
\label{eqn21}
\end{equation}%
where the eigenprojectors defined in Liouville space are 
\begin{equation}
P_{\nu }\equiv \left| \nu \right) \left( \nu \right| =\left| \left| j\left(
i\right) \otimes n\left( k\right) \right\rangle \left\langle j^{\prime
}\left( i\right) \otimes n^{\prime }\left( k\right) \right| \right) \left(
\left| j^{\prime }\left( i\right) \otimes n^{\prime }\left( k\right)
\right\rangle \left\langle j\left( i\right) \otimes n\left( k\right) \right|
\right| .  \label{eqn22}
\end{equation}%
Thus Eq.(\ref{eqn18}) remains invariant with or without the diagonal
interaction; that is, there is no decoherence introduced by the diagonal
interaction for the stationary states. But since the eigenvalues have been
changed by the interaction, the evolution of the density operators are still
subject to a type of ``decoherence'' (unitary error) introduced by this
change. That is, before the interaction the evolution is 
\begin{equation}
\left| \psi \left( t\right) \right\rangle _{b}=\sum_{jn}e^{-i\left( \left(
-1\right) ^{j}\omega _{0}+n\omega \right) t}\left\langle j\left( i\right)
\otimes n\left( k\right) \right. \left| \psi \left( t\right) \right\rangle
\left| j\left( i\right) \otimes n\left( k\right) \right\rangle ,
\label{eqn23}
\end{equation}%
while after interaction, the evolution is 
\begin{equation}
\left| \psi \left( t\right) \right\rangle _{a}=\sum_{jn}e^{-i\left( \left(
-1\right) ^{j}\omega _{0}+n\omega +\frac{1+\left( -1\right) ^{j}}{2}%
ng\right) t}\left\langle j\left( i\right) \otimes n\left( k\right) \right.
\left| \psi \left( t\right) \right\rangle \left| j\left( i\right) \otimes
n\left( k\right) \right\rangle ,  \label{eqn24}
\end{equation}%
and generally we have 
\begin{equation}
\left| \psi \left( t\right) \right\rangle _{b}\neq \left| \psi \left(
t\right) \right\rangle _{a}.  \label{eqn25}
\end{equation}%
In other words, Eqs.(\ref{eqn18}) and (\ref{eqn20}) reveal that the
eigenvectors of the total Hamiltonian $H$ with or without the diagonal
interaction, have the same tensor products, and are all diagonal with
respect to the set of these tensor products, showing that the diagonal
interaction does not introduce any decoherence in the stationary states.
Eqs.(\ref{eqn23}) and (\ref{eqn24}) show that the eigenvalues of the total
Hamiltonian are changed by the interaction. In this case, the interaction
introduces a phase shift which may introduce a type of ``decoherence''
(unitary error) in the evolution of the states, although the fidelity$^{%
\left[ 18\right] }$ $Tr\left( \sqrt{\rho \left( 0\right) \rho \left(
t\right) }\right) $, remains $1$ for both cases (i.e., with or without the
interaction).

\section{DF Subspace for Triangular Interaction}

The above discussion for a diagonal interaction, refers to a rather ideal
case. Another less ideal DF situation is now discussed - it is the
``triangular interaction''. This is defined as when the off-diagonal matrix
elements (of the interaction part of the total Hamiltonian) are up or lower
triangular. For example, if the interaction part of the Hamiltonian can be
expressed in a triangular form using the previous model 
\begin{equation}
H\left( 1,2\right) =g\left( a^{+}a\sigma ^{z}+a^{-}\sigma ^{-}\right) ,
\label{eqn26}
\end{equation}%
then the subdynamical kinetic equation (\ref{eqn14}) gives

\begin{eqnarray}
&&i\frac{\partial }{\partial t}\rho _{\nu }\left( t\right) =\left[ P_{\nu
}L_{S}P_{\nu }\rho _{\nu }\left( t\right) +P_{\nu }L_{R}P_{\nu }+P_{\nu
}L_{1}Q_{\nu }C_{\nu }\left( t\right) P_{\nu }\right] \rho _{\nu }\left(
t\right)  \nonumber \\
&=&P_{\nu }\left( L_{S}+L_{R}\right) P_{\nu }\rho _{\nu }\left( t\right) , 
\nonumber
\end{eqnarray}%
where from the triangular property of the matrix elements we have 
\begin{equation}
\left\langle i\otimes k\right| H\left( 1,2\right) \left| j\otimes
n\right\rangle =\left\{ 
\begin{array}{c}
\left( -1\right) ^{j}gn,\quad i=j,k=n \\ 
g\sqrt{n-1},\quad k=n-1,i=j-1 \\ 
0,\quad \text{otherwise}%
\end{array}%
\right. .  \label{eqn27}
\end{equation}%
This causes the second term in Eq.(\ref{eqn14}) to be zero, since the $%
C_{\nu }$ is a function of $Q_{\nu }L_{1}P_{\nu }$ from subdynamics theory$^{%
\left[ 6,7\right] }$: 
\begin{equation}
P_{\nu }L_{1}Q_{\nu }C_{\nu }P_{\nu }=0.  \label{eqn28}
\end{equation}%
Eqs.(\ref{eqn27}) and (\ref{eqn28}) show that the density operator $\rho
_{\nu }\left( t\right) $ is DF under the triangular interaction in the
subspaces. That is, with or without loading the triangular interaction, the
spectral decomposition for the intermediate operator $\Theta $ is invariant, 
\begin{equation}
\Theta =\sum_{\nu }P_{\nu }L_{0}P_{\nu }=\sum_{n\neq n^{\prime }}\sum_{j\neq
j^{\prime }=1}^{2}\left( -1\right) ^{j}ng\left| jn,j^{\prime }n^{\prime
}\right) \left( j^{\prime }n^{\prime },j,n\right| ,  \label{eqn29}
\end{equation}%
so that an arbitrary expanded projected density operator and its evolution
remains invariant with or without the interaction, 
\begin{equation}
\left| \rho ^{proj}\right) =\sum_{n\neq n^{\prime }}\sum_{j\neq j^{\prime
}=1}^{2}\left| \rho _{\nu }^{proj}\right) =\sum_{n\neq n^{\prime
}}\sum_{j\neq j^{\prime }=1}^{2}\left( j^{\prime }n^{\prime },j,n\right|
\left. \rho _{\nu }^{proj}\right) \left| jn,j^{\prime }n^{\prime }\right) ,
\label{eqn30}
\end{equation}%
and 
\begin{equation}
\left| \rho ^{proj}\left( t\right) \right) =\sum_{n\neq n^{\prime
}}\sum_{j\neq j^{\prime }=1}^{2}\left| \rho _{\nu }^{proj}\left( t\right)
\right) =\sum_{n\neq n^{\prime }}\sum_{j\neq j^{\prime }=1}^{2}e^{-i\left(
-1\right) ^{j}ngt}\left( j^{\prime }n^{\prime },j,n\right| \left. \rho _{\nu
}^{proj}\left( t\right) \right) \left| jn,j^{\prime }n^{\prime }\right) .
\label{eqn31}
\end{equation}%
This shows that no decoherence is introduced by the triangular interaction.
However, this does not mean that the total density operator $\rho \left(
t\right) $ is DF in the total space. Indeed, taking into account Eq.(\ref%
{eqn13}) and using the similarity operator $\Omega $ on Eq.(\ref{eqn14}),
gives\qquad 
\begin{equation}
i\frac{\partial }{\partial t}\Omega \rho _{\nu }\left( t\right) =i\frac{%
\partial }{\partial t}\rho \left( t\right) =\Omega \Theta _{\nu }\Omega
^{-1}\Omega \rho _{\nu }\left( t\right) =\Omega \Theta _{\nu }\Omega
^{-1}\rho \left( t\right) ,  \label{eqn32}
\end{equation}%
and yields 
\begin{eqnarray}
L &=&\sum_{\nu }\Omega \Theta _{\nu }\Omega ^{-1}  \label{eqn33} \\
&=&%
\mathrel{\mathop{\sum }\limits_{\nu }}%
\left( \nu \right| L_{0}\left| \nu \right) \left( P_{\nu }+C_{\nu }\right)
\left| \nu \right) \left( \nu \right| \left( P_{\nu }+D_{\nu }\right) 
\nonumber \\
&=&\sum_{n\neq n^{\prime }}%
\mathrel{\mathop{\sum^{2}}\limits_{j=1}}%
\left( -1\right) ^{j}g\left( n-n^{\prime }\right) \left( \left| \nu \right)
+\sum_{\nu ^{\prime }<\nu }\left( \nu ^{\prime }\right| C_{\nu }\left| \nu
\right) \left| \nu ^{\prime }\right) \right) \left( \left( \nu \right|
+\sum_{\nu ^{\prime }<\nu }\left( \nu \right| D_{\nu }\left| \nu ^{\prime
}\right) \left( \nu ^{\prime }\right| \right) ,  \nonumber
\end{eqnarray}%
where $\left( \nu ^{\prime }\right| C_{\nu }\left| \nu \right) $ can be
obtained from the recurrence formula (\ref{eqn6}) or (\ref{eqn9}), 
\begin{eqnarray}
&&\left( \nu ^{\prime }\right| C_{\nu }\left| \nu \right)  \label{eqn34} \\
&=&\frac{1}{l_{\nu }-l_{\nu ^{\prime }}}g\left( \nu ^{\prime }\right|
L_{1}\left( P_{\nu }+C_{\nu }\right) \left| \nu \right)  \nonumber \\
&=&\frac{1}{\left( \left( -1\right) ^{j}n-\left( -1\right) ^{j^{\prime
}}n^{\prime }\right) -\left( \left( -1\right) ^{i}m-\left( -1\right)
^{i^{\prime }}m^{\prime }\right) }  \nonumber \\
&&\times \left[ \left( im,i^{\prime }m^{\prime }\right| L_{1}\left|
jn,j^{\prime }n^{\prime }\right) +\left( 1m,2m^{\prime }\right| L_{1}\left|
2\left( m+1\right) ,1\left( m^{\prime }+1\right) \right) \left(
2l,1l^{\prime }\right| C_{jn,j^{\prime }n^{\prime }}\left| jn,j^{\prime
}n^{\prime }\right) \right]  \nonumber
\end{eqnarray}%
with 
\begin{eqnarray}
&&\left( im,i^{\prime }m^{\prime }\right| L_{1}\left| kl,k^{\prime
}l^{\prime }\right)  \label{eqn35} \\
&=&\left\langle im\right| H_{1}\left| kl\right\rangle \delta _{k^{\prime
}i^{\prime }}\delta _{l^{\prime }m^{\prime }}-\left\langle k^{\prime
}l^{\prime }\right| H_{1}\left| i^{\prime }m^{\prime }\right\rangle \delta
_{ik}\delta _{ml}  \nonumber \\
&=&g\sqrt{m-1}\delta _{i\left( k-1\right) }\delta _{m\left( l-1\right)
}\delta _{k^{\prime }i^{\prime }}\delta _{l^{\prime }m^{\prime }}-g\sqrt{%
l^{\prime }-1}\delta _{k^{\prime }\left( i^{\prime }-1\right) }\delta
_{l^{\prime }\left( m^{\prime }-1\right) }\delta _{ik}\delta _{ml}. 
\nonumber
\end{eqnarray}%
Hence the eigenvectors of the total Hamiltonian can be formally written as 
\begin{equation}
\left| \varphi _{\nu }\right) =\left| \nu \right) +\sum_{\nu ^{\prime }<\nu
}\left( \nu ^{\prime }\right| C_{\nu }\left( t\right) \left| \nu \right)
\left| \nu ^{\prime }\right) ,  \label{eqn36}
\end{equation}%
and 
\begin{equation}
\left( \varphi _{\nu }\right| =\left( \nu \right| +\sum_{\nu ^{\prime }<\nu
}\left( \nu \right| D_{\nu }\left( t\right) \left| \nu ^{\prime }\right)
\left( \nu ^{\prime }\right| .  \label{eqn37}
\end{equation}%
This shows that the eigenvectors of the total Hamiltonian are a combination
of the eigenvectors of the intermediate operator $\Theta $, and the total
system\ is still subject to decoherence. Therefore, it can be concluded that
the system, under a triangular interaction in each projected $P_{\nu }$
subspace, is DF for both stationary and evolutionary states, but suffers
from decoherence in the total space.

\section{DF and Decoherent Subspaces for a General Interaction}

Since the triangular interaction does not introduce decoherence in the
corresponding subdynamical projected subspaces, this suggests that there
should exist a more general class of DF subspaces. In fact, as mentioned in
the introduction, the existence of such DF subspaces has been shown by
several approaches, such as projection onto a symmetric subspace of multiple
copies of a quantum computing system$^{\left[ 4\right] }$. In this
treatment, we focus on providing an argument based on a subdynamics
approach. Generally, it is assumed that the interaction parts of the
Hamiltonian for the previous model are non-diagonal or non-triangular in
form; moreover, for practical use (i.e., to construct a quantum logical
gate), a model is considered consisting of two atoms with two-levels
simultaneously interacting with a single optical mode$^{\left[ 19,20\right]
} $ and influenced by an environment consisting of a set of infinite
Harmonic oscillators. The Hamiltonian is given by 
\begin{equation}
H=H_{0}+\lambda H_{I}  \label{eqn38}
\end{equation}%
with 
\begin{eqnarray}
H_{0} &=&\sum_{j=1}^{2}\left[ \omega _{j}\sigma _{j}^{z}+g\left( a^{+}\sigma
_{j}^{-}+a\sigma _{j}^{+}\right) \right] +\omega a^{+}a+\sum_{k}\omega
_{k}b_{k}^{+}b_{k},  \label{eqn39} \\
H_{I} &=&\sum_{k}\sum_{j=1}^{2}g_{k}\left( b_{k}^{+}+b_{k}\right) \left(
\sigma _{j}^{-}+\sigma _{j}^{+}\right) ,  \label{eqn40}
\end{eqnarray}%
where $b_{k}^{+}$, $b_{k}$ are creation and annihilation operators of the
oscillator. The eigenvalues of the equation for $H_{0}$ are given by first
writing the matrix expression of $H_{0}$ in Hilbert space $\left\{ \left|
\pm \left( 1\right) ,\pm \left( 2\right) ,n,\left\{ n_{k}\right\}
\right\rangle \right\} $ and then diagonalizing it. Using the matrix
elements $\left\langle \pm \left( 1\right) ,\pm \left( 2\right) ,n^{\prime
},\left\{ n_{k}^{\prime }\right\} \right| H_{0}\left| \pm \left( 1\right)
,\pm \left( 2\right) ,n,\left\{ n_{k}\right\} \right\rangle $, the matrix
expression for $H_{0}$ in the subspace $\left\{ \varphi _{n,k}^{++},\varphi
_{n+1,k}^{-+},\varphi _{n+1,k}^{+-}\right\} ,$%
\begin{eqnarray}
\left| \varphi _{n,k}^{++}\right\rangle \bigskip &=&\left|
+(1),+(2),n,\left\{ n_{k}\right\} \right\rangle ,  \label{eqn41} \\
\left| \varphi _{n+1,k}^{-+}\right\rangle \bigskip &=&\left|
-(1),+(2),n+1,\left\{ n_{k}\right\} \right\rangle ,  \nonumber \\
\left| \varphi _{n+1,k}^{+-}\right\rangle \bigskip &=&\left|
+(1),-(2),n+1,\left\{ n_{k}\right\} \right\rangle ,  \nonumber
\end{eqnarray}%
is given by 
\begin{equation}
H_{0}=\left( 
\begin{array}{lll}
\frac{\omega _{1}+\omega _{2}}{2}+n\omega +\sum_{k}n_{k}\omega _{k} & g\sqrt{%
n+1} & g\sqrt{n+1} \\ 
g\sqrt{n+1} & (n+1)\omega +\sum_{k}n_{k}\omega _{k} & 0 \\ 
g\sqrt{n+1} & 0 & (n+1)\omega +\sum_{k}n_{k}\omega _{k}%
\end{array}%
\right) .  \label{eqn42bb}
\end{equation}%
Following the eigenvalue equation, $H_{0}\left| f\right\rangle =\varepsilon
\left| f\right\rangle $, gives a characteristic equation, 
\begin{equation}
\left\| 
\begin{array}{lll}
a & \gamma & \gamma \\ 
\gamma & b & 0 \\ 
\gamma & 0 & b%
\end{array}%
\right\| =0,  \label{eqn43}
\end{equation}%
its solutions give the eigenvalues as: 
\begin{eqnarray}
\varepsilon _{n,k}^{++} &=&b,  \label{eqn44} \\
\varepsilon _{n+1,k}^{-+} &=&\frac{1}{2}b+\frac{1}{2}a+\frac{1}{2}\sqrt{%
\left( b^{2}-2ab+a^{2}+8\gamma ^{2}\right) },  \nonumber \\
\varepsilon _{n+1,k}^{+-} &=&\frac{1}{2}b+\frac{1}{2}\allowbreak a-\frac{1}{2%
}\sqrt{\left( b^{2}-2ab+a^{2}+8\gamma ^{2}\right) },  \nonumber
\end{eqnarray}%
and the corresponding eigenvectors are 
\begin{equation}
\left| f_{n,k}^{++}\right\rangle =-\left| \varphi _{n+1,k}^{-+}\right\rangle
+\left| \varphi _{n+1,k}^{+-}\right\rangle ,  \label{eqn45}
\end{equation}%
\[
\left| f_{n+1,k}^{-+}\right\rangle =-\frac{\frac{1}{2}b-\frac{1}{2}a+\frac{1%
}{2}\sqrt{\left( b^{2}-2ab+a^{2}+8\gamma ^{2}\right) }}{\gamma }\left|
\varphi _{n,k}^{++}\right\rangle +\left| \varphi _{n+1,k}^{-+}\right\rangle
+\left| \varphi _{n+1,k}^{+-}\right\rangle , 
\]%
\[
\left| f_{n+1,k}^{+-}\right\rangle =-\frac{\frac{1}{2}b-\frac{1}{2}a-\frac{1%
}{2}\sqrt{\left( b^{2}-2ab+a^{2}+8\gamma ^{2}\right) }}{\gamma }\left|
\varphi _{n,k}^{++}\right\rangle +\left| \varphi _{n+1,k}^{-+}\right\rangle
+\left| \varphi _{n+1,k}^{+-}\right\rangle , 
\]%
where we denote $a$ $\equiv $ $\frac{\omega _{1}+\omega _{2}}{2}$ + $n\omega 
$ + $\sum_{k}n_{k}\omega _{k}-\varepsilon $, $b$ $\equiv $ $\left(
n+1\right) \omega $ + $\sum_{k}n_{k}\omega _{k}-\varepsilon $ and $\gamma $ $%
\equiv \sqrt{g+1}$.

From these eigenvectors, the eigenprojectors $P_{nm,kj}^{\pm \mp }$ of the
corresponding unperturbed Liouvillian $L_{0}$ are defined as 
\begin{eqnarray}
P_{nm,kj}^{\pm \mp } &=&\left| \left| f_{n,k}^{\pm }\right\rangle
\left\langle f_{m,j}^{\mp }\right| \right) \left( \left| f_{m,j}^{\mp
}\right\rangle \left\langle f_{n,k}^{\pm }\right| \right|  \label{eqn46} \\
&=&\left| \phi _{nm,jk}^{\pm \mp }\right) \left( \phi _{nm,jk}^{\pm \mp
}\right|  \nonumber
\end{eqnarray}%
with 
\begin{equation}
Q_{nm,kj}^{\pm \mp }=1-P_{nm,kj}^{\pm \mp }.  \label{eqn47}
\end{equation}%
Without the interaction, the spectral decomposition for $L_{0}$ is given by 
\begin{eqnarray}
L_{0} &=&\sum_{\pm ,\mp }\sum_{nm,kj}\left( \varepsilon _{n,k}^{\pm
}-\varepsilon _{m,j}^{\pm }\right) \left| \phi _{nm,jk}^{\pm \mp }\right)
\left( \phi _{nm,jk}^{\pm \mp }\right|  \label{eqn48} \\
&=&\sum_{\pm ,\mp }\sum_{nm,kj}\left( \varepsilon _{n,k}^{\pm }-\varepsilon
_{m,j}^{\pm }\right) P_{nm,kj}^{\pm \mp }.  \nonumber
\end{eqnarray}%
The corresponding spectral decomposition of the intermediate operator $%
\Theta $, in terms of subdynamics theory, is the same as that for $L_{0}$,
i.e., 
\begin{equation}
\Theta =\sum_{\pm ,\mp }\sum_{nm,kj}P_{nm,kj}^{\pm \mp }L_{0}P_{nm,kj}^{\pm
\mp }=\sum_{\pm ,\mp }\sum_{nm,kj}\left( \varepsilon _{n,k}^{\pm
}-\varepsilon _{m,j}^{\pm }\right) \left| \phi _{nm,jk}^{\pm \mp }\right)
\left( \phi _{nm,jk}^{\pm \mp }\right| .  \label{eqn49}
\end{equation}%
Now considering interactions occurring to this system, the Liouvillian
becomes 
\begin{equation}
L=L_{0}+\lambda L_{1}  \label{eqn50}
\end{equation}%
with 
\begin{equation}
L_{1}=\sum_{nm,jk\neq n^{\prime }m^{\prime },j^{\prime }k^{\prime }}\left(
\phi _{nm,jk}^{\pm \mp }\right| L_{I}\left| \phi _{n^{\prime }m^{\prime
},j^{\prime }k^{\prime }}^{\mp \pm }\right) \left| \phi _{nm,jk}^{\pm \mp
}\right) \left( \phi _{n^{\prime }m^{\prime },j^{\prime }k^{\prime }}^{\mp
\pm }\right| .  \label{eqn51}
\end{equation}%
Then, in terms of the subdynamical equation (\ref{eqn3aa}) the operator $%
\Theta $ can be expressed as%
\begin{equation}
\Theta =\sum_{\pm ,\mp }\sum_{nm,kj}\left( P_{nm,kj}^{\pm \mp
}L_{0}P_{nm,kj}^{\pm \mp }+P_{nm,kj}^{\pm \mp }L_{1}Q_{nm,kj}^{\pm \mp
}C_{nm,jk}^{\pm \mp }P_{nm,kj}^{\pm \mp }\right) ,  \label{eqn3ab}
\end{equation}%
and the corresponding spectral decomposition is 
\begin{eqnarray}
&&\Theta  \label{eqn52} \\
&=&\sum_{\pm ,\mp }\sum_{nm,kj}\left\{ \left( \varepsilon _{n,k}^{\pm
}-\varepsilon _{m,j}^{\pm }\right) +\left( \phi _{nm,jk}^{\pm \mp }\right|
L_{1}C_{nm,jk}^{\pm \mp }\left| \phi _{nm,jk}^{\pm \mp }\right) \right\}
\left| \phi _{nm,jk}^{\pm \mp }\right) \left( \phi _{nm,jk}^{\pm \mp
}\right| .  \nonumber
\end{eqnarray}%
This again means that the eigenvectors of $\Theta $, $\phi _{nm,jk}^{\pm \mp
}$, with or without coupling are the same, while the corresponding
eigenvalues change to $\left\{ \left( \varepsilon _{n,k}^{\pm }-\varepsilon
_{m,j}^{\pm }\right) +\left( \phi _{nm,jk}^{\pm \mp }\right|
L_{1}C_{nm,jk}^{\pm \mp }\left| \phi _{nm,jk}^{\pm \mp }\right) \right\} $
(with the interaction) from $\left( \varepsilon _{n,k}^{\pm }-\varepsilon
_{m,j}^{\pm }\right) $ (without the interaction). Using the set of the
eigenvectors $\left\{ \left| \phi _{nm,jk}^{\pm \mp }\right) \text{, }\left(
\phi _{nm,jk}^{\pm \mp }\right| \right\} $ of the free Liouvillian $L_{0}$
as a basis for expanding an arbitrary state of the system $\rho ^{proj}$, in
the projected subspace space, the stationary projected density operator
remains invariant with or without the interaction, 
\begin{equation}
\left| \rho ^{proj}\right) \equiv \sum_{\pm }\sum_{nm,kj}\left|
P_{nm,kj}^{\pm \mp }\Pi _{nm,kj}^{\pm \mp }\rho \right) =\sum_{\pm
}\sum_{nm,kj}\left( \phi _{nm,jk}^{\pm \mp }\right. \left| P_{nm,kj}^{\pm
\mp }\Pi _{nm,kj}^{\pm \mp }\rho \right) \left| \phi _{nm,jk}^{\pm \mp
}\right) ,  \label{eqn53}
\end{equation}%
while the evolution of the projected density operator changes, upon loading
the interaction, to 
\begin{eqnarray}
\left| \rho ^{proj}\left( t\right) \right) &\equiv &\sum_{\pm
}\sum_{nm,kj}\left| P_{nm,kj}^{\pm \mp }\Pi _{nm,kj}^{\pm \mp }\rho \left(
t\right) \right)  \label{eqn54} \\
&=&\sum_{\pm }\sum_{nm,kj}e^{-i\left\{ \left( \varepsilon _{n,k}^{\pm
}-\varepsilon _{m,j}^{\pm }\right) +\left( \phi _{nm,jk}^{\pm \mp }\right|
L_{1}C_{nm,jk}^{\pm \mp }\left| \phi _{nm,jk}^{\pm \mp }\right) \right\}
t}\left( \phi _{nm,jk}^{\pm \mp }\right. \left| P_{nm,kj}^{\pm \mp }\Pi
_{nm,kj}^{\pm \mp }\rho \left( 0\right) \right) \left| \phi _{nm,jk}^{\pm
\mp }\right)  \nonumber \\
&\neq &\sum_{\pm }\sum_{nm,kj}e^{-i\left( \varepsilon _{n,k}^{\pm
}-\varepsilon _{m,j}^{\pm }\right) t}\left( \phi _{nm,jk}^{\pm \mp }\right.
\left| P_{nm,kj}^{\pm \mp }\Pi _{nm,kj}^{\pm \mp }\rho \left( 0\right)
\right) \left| \phi _{nm,jk}^{\pm \mp }\right) .  \nonumber
\end{eqnarray}%
This is similar to the case for the diagonal interaction, but it takes place
in the projected subspace, showing that no decoherence occurs to the
stationary states, but the evolutionary states are subject to a phase shift
introduced by the change in the eigenvalues in the subdynamical projected
subspace. This phase shift induces a type of ``decoherence'' (like a unitary
error) to the evolutionary states. Even so, the fidelity of the mixed states
in the subspace can be shown to equal $1$,%
\begin{eqnarray}
&&F\left( t\right)  \label{eqn3ag} \\
&=&Tr\sqrt{\rho ^{proj}\left( 0\right) \rho ^{proj}\left( t\right) } 
\nonumber \\
&=&Tr\sqrt{\rho ^{proj}\left( 0\right) \widehat{T}e^{-i\int \Theta \left(
t^{\prime }\right) dt^{\prime }}\rho ^{proj}\left( 0\right) \widehat{T}%
e^{i\int \Theta \left( t^{\prime }\right) dt^{\prime }}}  \nonumber \\
&=&Tr\sqrt{\left( \sum_{\pm }\sum_{nm,kj}\left( \rho ^{proj}\right)
_{nm,kj}^{\pm \mp }P_{nm,kj}^{\pm \mp }\right) \left( \widehat{T}e^{-i\int
\Theta \left( t^{\prime }\right) dt^{\prime }}\sum_{\pm }\sum_{nm,kj}\left(
\rho ^{proj}\right) _{nm,kj}^{\pm \mp }P_{nm,kj}^{\pm \mp }e^{i\int \Theta
\left( t^{\prime }\right) dt^{\prime }}\right) }  \nonumber \\
&=&Tr\left\{ \sqrt{\sum_{\pm }\sum_{nm,kj}\left[ \left( \rho ^{proj}\right)
_{nm,kj}^{\pm \mp }\right] ^{2}e^{-i\left[ \left( \varepsilon _{n,k}^{\pm
}-\varepsilon _{m,j}^{\pm }\right) +\left( \phi _{nm,jk}^{\pm \mp }\right|
L_{1}C_{nm,jk}^{\pm \mp }\left| \phi _{nm,jk}^{\pm \mp }\right) \right] t}}%
\right.  \nonumber \\
&&\times \left. \sqrt{P_{nm,kj}^{\pm \mp }e^{i\left[ \left( \varepsilon
_{n,k}^{\pm }-\varepsilon _{m,j}^{\pm }\right) +\left( \phi _{nm,jk}^{\pm
\mp }\right| L_{1}C_{nm,jk}^{\pm \mp }\left| \phi _{nm,jk}^{\pm \mp }\right) %
\right] t}}\right\}  \nonumber \\
&=&Tr\sqrt{\sum_{\pm }\sum_{nm,kj}\left[ \left( \rho ^{proj}\right)
_{nm,kj}^{\pm \mp }\right] ^{2}P_{nm,kj}^{\pm \mp }}  \nonumber \\
&=&Tr\left( \sqrt{\rho ^{proj}\left( t_{0}\right) \rho ^{proj}\left(
t_{0}\right) }\right) =1,  \nonumber
\end{eqnarray}%
where $\left( \rho ^{proj}\right) _{nm,kj}^{\pm \mp }$ is the matrix element
of the density operator $\rho ^{proj}\left( t_{0}\right) $ with respect to $%
\left| \phi _{nm,jk}^{\pm \mp }\right) $, $\left( \phi _{nm,jk}^{\pm \mp
}\right| $, and $\rho ^{proj}\left( t_{0}\right) $ can be expanded by the
set of the eigenprojectors $P_{nm,kj}^{\pm \mp }$. Since the eigenprojectors
remain invariant, the initial density operator remains the same with or
without the interaction under the assumption of decoupling of initial
states. Eq.(\ref{eqn3ag}) shows that the constructed projected subspace is
DF! Whatever the subspace, the total density operator $\rho $ is decoherent
in the total space. This suggests a universal procedure for constructing a
DF subdynamical projected subspace by choosing a set of one-dimensional
eigenprojectors $P_{\nu }$ of the unperturbed Liouville operator $%
L_{0}\left( t\right) $ through the kinetic equation (\ref{eqn14}) or (\ref%
{eqn3aa}). In this constructed subspace, the governing equation is the
subdynamical kinetic equation and the interaction part of the total
Liouvillian cannot change the diagonal property of the intermediate
operator, if each of $P_{\nu }$ in the complete set is chosen as a
one-dimensional projector by construction (note if $P_{\nu }$ is over
one-dimensional the intermediate operator generally is not diagonal).

The above discussions are summarized below in tabular form:

\[
\begin{tabular}{|l|ll|ll|}
\hline
$Interaction\text{ }Type$ & $Total$ & $Space$ & $\Pr ojected$ & $Subspace$
\\ \hline
& $St$ & \multicolumn{1}{|l|}{$Et$} & $St$ & \multicolumn{1}{|l|}{$Et$} \\ 
\hline
$\text{Diagonal Interaction}$ & $DF$ & \multicolumn{1}{|l|}{$DF$} & $DF$ & 
\multicolumn{1}{|l|}{$PE$} \\ \hline
$\text{Triangular Interaction}$ & $DF$ & \multicolumn{1}{|l|}{$DF$} & $DF$ & 
\multicolumn{1}{|l|}{$DF$} \\ \hline
$\text{Diagonal }\Theta $ & $D$ & \multicolumn{1}{|l|}{$D$} & $DF$ & 
\multicolumn{1}{|l|}{$PE$} \\ \hline
\end{tabular}%
\]%
where $St$ and $Et$ refer to stationary states and evolutionary states, and $%
PE$, $D$ and $DF$ refer to a phase error, decoherent and decoherence-free,
respectively. The table shows that it is possible to construct a DF
subdynamical projected subspace for an open quantum computing system with
three types of interaction occurring with the environment. This allows
quantum logical computations to be performed by using the projected density
operator on the corresponding subspaces, which may be beyond Hilbert space.
For example, these can include rigged Hilbert spaces. To clarify this point,
a model of two two-level atoms is discussed in detail in the following
subsection. A second-order calculation is presented, but the approximation
does not restrict the validity of the result, since there is no problem in
allowing high order calculations to be performed based on the same
methodology.

\section{Quantum Logical Operation in Subspaces}

The expected evolution of the projected density operator, in the projected
subspaces, can be formally solved from Eq.(\ref{eqn3aa}), 
\begin{eqnarray}
\left| \rho ^{proj}\left( t\right) \right) &=&\sum_{\pm \mp
}\sum_{nmjk}e^{-i\Theta _{nm,kj}^{\pm \pm \mp \mp }t}\left| P_{nm,kj}^{\pm
\pm \mp \mp }\Pi _{nm,kj}^{\pm \pm \mp \mp }\rho \left( 0\right) \right)
\label{eqn55aa} \\
&=&\sum_{\pm \mp }\sum_{nmjk}e^{-iE_{nm,kj}^{\pm \pm \mp \mp }t}\left( \phi
_{nm,jk}^{\pm \pm \mp \mp }\right. \left| \rho _{nm,kj}^{\pm \pm \mp \mp
}\left( 0\right) \right) \left| \phi _{nm,jk}^{\pm \pm \mp \mp }\right) , 
\nonumber
\end{eqnarray}%
where the difference in energy $E_{nm,kj}^{\pm \pm \mp \mp }$ is defined as $%
E_{nm,kj}^{0}$ + $\triangle E_{nm,kj}$ = $\varepsilon _{n,k}^{\pm \pm
}-\varepsilon _{m,j}^{\mp \mp }$ + $\lambda \left( \phi _{nm,jk}^{\pm \pm
\mp \mp }\right| L_{1}C_{nm,jk}^{\pm \pm \mp \mp }\left| \phi _{nm,jk}^{\pm
\pm \mp \mp }\right) $, and the projected density operator is determined
from $\rho \left( t\right) $, 
\begin{eqnarray}
P_{nm,kj}^{\pm \pm \mp \mp }\Pi _{nm,kj}^{\pm \pm \mp \mp }\rho \left(
t\right) &\equiv &\rho _{nm,kj}^{\pm \pm \mp \mp }\left( t\right)
\label{eqn55} \\
&=&\left( P_{nm,kj}^{\pm \pm \mp \mp }+D_{nm,kj}^{\pm \pm \mp \mp
}C_{nm,kj}^{\pm \pm \mp \mp }\right) ^{-1}\left( P_{nm,kj}^{\pm \pm \mp \mp
}+D_{nm,kj}^{\pm \pm \mp \mp }\right) \rho \left( t\right) .  \nonumber
\end{eqnarray}%
Here $D_{nm,kj}^{\pm \pm \mp \mp }$ is given by Eq.(\ref{eqn10}), 
\begin{eqnarray}
&&D_{nm,kj}^{\pm \pm \mp \mp }  \label{eqn56} \\
&=&-i\lambda e^{iP_{nm,jk}^{\pm \pm \mp \mp }LP_{nm,jk}^{\pm \pm \mp \mp }t} 
\left[ \int_{\mp \infty }^{t}e^{-iP_{nm,jk}^{\pm \pm \mp \mp
}LP_{nm,jk}^{\pm \pm \mp \mp }\tau }P_{nm,jk}^{\pm \pm \mp \mp
}L_{1}Q_{nm,jk}^{\pm \pm \mp \mp }e^{iQ_{nm,jk}^{\pm \pm \mp \mp
}LQ_{nm,jk}^{\pm \pm \mp \mp }\tau }d\tau \right]  \nonumber \\
&&\times e^{-iQ_{nm,jk}^{\pm \pm \mp \mp }LQ_{nm,jk}^{\pm \pm \mp \mp }t} 
\nonumber \\
&=&-\lambda P_{nm,jk}^{\pm \pm \mp \mp }L_{1}Q_{nm,jk}^{\pm \pm \mp \mp }%
\frac{1}{Q_{nm,jk}^{\pm \pm \mp \mp }LQ_{nm,jk}^{\pm \pm \mp \mp }-\left(
\varepsilon _{n,k}^{\pm \pm }-\varepsilon _{m,j}^{\mp \mp }\right) }. 
\nonumber
\end{eqnarray}%
Taking into account the Born approximation, the second order term $%
D_{nm,jk}^{\left[ 2\right] \pm \pm \mp \mp }$ is 
\begin{eqnarray}
&&D_{nm,jk}^{\left[ 2\right] \pm \pm \mp \mp }  \label{eqn57} \\
&=&\sum_{\pm \mp }\sum_{n^{\prime \prime }m^{\prime \prime }j^{\prime \prime
}k^{\prime \prime }}\frac{-\lambda }{\left( \varepsilon _{n^{\prime
},k}^{\pm \pm }-\varepsilon _{m^{\prime },j}^{\mp \mp }\right) -\left(
\varepsilon _{n,k}^{\pm \pm }-\varepsilon _{m,j}^{\mp \mp }\right) \pm i0}%
\left[ \left( \phi _{nm,jk}^{\pm \pm \mp \mp }\right| L_{1}\left| \phi
_{n^{\prime \prime }m^{\prime \prime },j^{\prime \prime }k^{\prime \prime
}}^{\pm \pm \mp \mp }\right) \right.  \nonumber \\
&&\left. -\frac{\lambda \left( \phi _{nm,jk}^{\pm \pm \mp \mp }\right|
L_{1}^{2}\left| \phi _{n^{\prime \prime }m^{\prime \prime },j^{\prime \prime
}k^{\prime \prime }}^{\pm \pm \mp \mp }\right) }{\left( \varepsilon
_{n^{\prime \prime },k^{\prime \prime }}^{\pm \pm }-\varepsilon _{m^{\prime
\prime }j^{\prime \prime }}^{\mp \mp }\right) -\left( \varepsilon
_{n,k}^{\pm \pm }-\varepsilon _{m,j}^{\mp \mp }\right) \pm i0}\right] \left|
\phi _{nm,jk}^{\pm \pm \mp \mp }\right) \left( \phi _{n^{\prime \prime
}m^{\prime \prime },j^{\prime \prime }k^{\prime \prime }}^{\pm \pm \mp \mp
}\right| +0\left( \lambda ^{2}\right) ,  \nonumber
\end{eqnarray}%
which gives the second order projected density operator as%
\begin{equation}
\rho _{nm,kj}^{\left[ 2\right] \pm \pm \mp \mp }\left( t\right) =\left[
P_{nm,kj}^{\pm \pm \mp \mp }+D_{nm,kj}^{\left[ 2\right] \pm \pm \mp \mp }%
\right] \rho \left( t\right) ,  \label{eqn58}
\end{equation}%
where $\left( \phi _{nm,jk}^{\pm \pm \mp \mp }\right| L_{1}^{2}\left| \phi
_{nm,jk}^{\pm \pm \mp \mp }\right) $ = $g_{j}\left( \sqrt{\left( j+1\right) j%
}+\sqrt{\left( j-1\right) j}\right) -g_{k}\left( \sqrt{\left( k+1\right) k}+%
\sqrt{\left( k-1\right) k}\right) $.

On the other hand, $E_{nm,kj}^{\pm \pm \mp \mp }$ can be approximately found
by calculating $C_{nm,jk}^{\pm \pm \mp \mp }$ from Eq.(\ref{eqn9}). For
example, the first order $C_{nm,kj}^{\left[ 1\right] \pm \pm \mp \mp }$ is 
\begin{eqnarray}
&&C_{nm,jk}^{\left[ 1\right] \pm \pm \mp \mp }=-\lambda \frac{1}{%
Q_{nm,jk}^{\pm \pm \mp \mp }LQ_{nm,jk}^{\pm \pm \mp \mp }-\left( \varepsilon
_{n,k}^{\pm \pm }-\varepsilon _{m,j}^{\mp \mp }\right) }Q_{nm,jk}^{\pm \pm
\mp \mp }L_{1}P_{nm,jk}^{\pm \pm \mp \mp }  \label{eqn59} \\
&=&-\lambda \sum_{\pm \mp }\sum_{n^{\prime \prime }m^{\prime \prime
}j^{\prime \prime }k^{\prime \prime }}\frac{1}{\left( \varepsilon
_{n^{\prime \prime },k^{\prime \prime }}^{\pm \pm }-\varepsilon _{m^{\prime
\prime },j^{\prime \prime }}^{\mp \mp }\right) -\left( \varepsilon
_{n,k}^{\pm \pm }-\varepsilon _{m,j}^{\mp \mp }\right) \pm i0}  \nonumber \\
&&\times \left( \phi _{n^{\prime \prime }m^{\prime \prime },j^{\prime \prime
}k^{\prime \prime }}^{\pm \pm \mp \mp }\right| L_{1}\left| \phi
_{nm,jk}^{\pm \pm \mp \mp }\right) \left| \phi _{n^{\prime \prime }m^{\prime
\prime },j^{\prime \prime }k^{\prime \prime }}^{\pm \pm \mp \mp }\right)
\left( \phi _{nm,jk}^{\pm \pm \mp \mp }\right| +0\left( \lambda \right) , 
\nonumber
\end{eqnarray}%
which then gives the second order term $E_{nm,kj}^{\left[ 2\right] \pm \pm
\mp \mp }$, 
\begin{eqnarray}
&&E_{nm,kj}^{\left[ 2\right] \pm \pm \mp \mp }=\left( \varepsilon
_{n,k}^{\pm \pm }-\varepsilon _{m,j}^{\mp \mp }\right) -\lambda
^{2}\sum_{\pm \mp }\sum_{n^{\prime }m^{\prime },j^{\prime }k^{\prime }}\frac{%
\left( \phi _{nm,jk}^{\pm \pm \mp \mp }\right| L_{1}^{2}\left| \phi
_{nm,jk}^{\pm \pm \mp \mp }\right) }{\left( \varepsilon _{n^{\prime \prime
},k^{\prime \prime }}^{\pm \pm }-\varepsilon _{m^{\prime \prime },j^{\prime
\prime }}^{\mp \mp }\right) -\left( \varepsilon _{n,k}^{\pm \pm
}-\varepsilon _{m,j}^{\mp \mp }\right) \pm i0}  \label{eqn60} \\
&=&\left( \varepsilon _{n,k}^{\pm \pm }-\varepsilon _{m,j}^{\mp \mp }\right)
-\lambda \sum_{\pm \mp }\sum_{n^{\prime }m^{\prime },j^{\prime }k^{\prime }}%
\frac{g_{j}\left( \sqrt{\left( j+1\right) j}+\sqrt{\left( j-1\right) j}%
\right) -g_{k}\left( \sqrt{\left( k+1\right) k}+\sqrt{\left( k-1\right) k}%
\right) }{\left( \varepsilon _{n^{\prime \prime },k^{\prime \prime }}^{\pm
\pm }-\varepsilon _{m^{\prime \prime },j^{\prime \prime }}^{\mp \mp }\right)
-\left( \varepsilon _{n,k}^{\pm \pm }-\varepsilon _{m,j}^{\mp \mp }\right)
\pm i0}.  \nonumber
\end{eqnarray}%
and hence giving the second order evolution of $\rho ^{proj\left[ 2\right]
}\left( t\right) $, 
\begin{equation}
\left| \rho ^{proj\left[ 2\right] }\left( t\right) \right) =\sum_{\pm \mp
}\sum_{nmjk}\left| \rho _{nm,kj}^{\left[ 2\right] \pm \pm \mp \mp }\left(
t\right) \right) =\sum_{\pm \mp }\sum_{nmjk}e^{-iE_{nm,kj}^{\left[ 2\right]
\pm \pm \mp \mp }t}\left( \phi _{nm,jk}^{\pm \pm \mp \mp }\right. \left|
\rho _{nm,kj}^{\left[ 2\right] \pm \pm \mp \mp }\left( 0\right) \right)
\left| \phi _{nm,jk}^{\pm \pm \mp \mp }\right) .  \label{eqn3ac}
\end{equation}%
Considering this evolution, a quantum XOR operator for the system may be
constructed by applying a sequence of operations$^{\left[ 20,21\right] }$
which is related the swap operator, 
\begin{equation}
U_{sw}\left| i,j\right\rangle =e^{-iLt_{sw}}\left| i,j\right\rangle =\left|
j,i\right\rangle .  \label{eqn62}
\end{equation}%
Suppose there is no coupling interaction, the ideal swap operator is then
given by adjusting the coupling time $t_{sw}$ 
\begin{eqnarray}
U_{sw} &=&\sum_{\pm \mp }\sum_{nmjk}e^{-iE_{nk,mj}^{0}t_{sw}}  \label{eqn63}
\\
&=&\sum_{\pm \mp }\sum_{nmjk}e^{-i\left( \varepsilon _{n,k}^{\pm \pm
}-\varepsilon _{m,j}^{\mp \mp }\right) t_{sw}}\left| \phi _{nm,jk}^{\pm \pm
\mp \mp }\right) \left( \phi _{nm,jk}^{\pm \pm \mp \mp }\right| ,  \nonumber
\end{eqnarray}%
and loading the interaction, the nonideal action of the swap operator is
given by the evolution operator (since the energy shift induced an error),
i.e., 
\begin{eqnarray}
U_{sw} &=&\sum_{\pm \mp }\sum_{nmjk}e^{-i\left( E_{nk,mj}^{0}+\triangle
E_{nk,mj}\right) \left( t_{sw}+\triangle t\right) }\left| \phi _{nm,jk}^{\pm
\pm \mp \mp }\right) \left( \phi _{nm,jk}^{\pm \pm \mp \mp }\right|
\label{eqn64b} \\
&=&\sum_{\pm \mp }\sum_{nmjk}e^{-iE_{nk,mj}^{\pm \pm \mp \mp }\left(
t_{sw}+\triangle t\right) }\left| \phi _{nm,jk}^{\pm \pm \mp \mp }\right)
\left( \phi _{nm,jk}^{\pm \pm \mp \mp }\right| .  \nonumber
\end{eqnarray}%
To cancel the decoherence, we have to have%
\begin{equation}
e^{-i\left( \varepsilon _{n,k}^{\pm \pm }-\varepsilon _{m,j}^{\mp \mp
}\right) t_{sw}}=e^{-iE_{nm,kj}^{\pm \pm \mp \mp }\left( t_{sw}+\triangle
t_{sw}\right) },  \label{eqn65b}
\end{equation}%
which gives%
\begin{equation}
\triangle t_{sw}=\frac{t_{sw}}{\frac{E^{0}}{\triangle E}+1},  \label{eqn65c}
\end{equation}%
where we have assumed that the distribution of energy is homogeneous, $\frac{%
E^{0}}{\triangle E}=$ constant, to determine the universal time $\triangle t$%
. For example, if the second order nonideal swap operator is 
\begin{eqnarray}
U_{sw}^{\left[ 2\right] } &=&\sum_{\pm \mp }\sum_{nmjk}e^{-i\left\{ \left(
\varepsilon _{n,k}^{\pm \pm }-\varepsilon _{m,j}^{\mp \mp }\right) -\lambda
\sum_{\pm \mp }\sum_{n^{\prime }m^{\prime },j^{\prime }k^{\prime }}\frac{%
g_{j}\left( \sqrt{\left( j+1\right) j}+\sqrt{\left( j-1\right) j}\right)
-g_{k}\left( \sqrt{\left( k+1\right) k}+\sqrt{\left( k-1\right) k}\right) }{%
\left( \varepsilon _{n^{\prime },k}^{\pm \pm }-\varepsilon _{m^{\prime
},j}^{\mp \mp }\right) -\left( \varepsilon _{n,k}^{\pm \pm }-\varepsilon
_{m,j}^{\mp \mp }\right) \pm i0}\right\} \left( t_{sw}+\triangle t_{sw}^{%
\left[ 2\right] }\right) }  \label{eqn66gg} \\
&&\times \left| \phi _{nm,jk}^{\pm \pm \mp \mp }\right) \left( \phi
_{nm,jk}^{\pm \pm \mp \mp }\right| .  \nonumber
\end{eqnarray}%
then the second order $\triangle t_{sw}^{\left[ 2\right] }$ is 
\begin{eqnarray}
&&\triangle t_{sw}^{\left[ 2\right] }  \label{eqn55s} \\
&=&-\frac{t_{sw}\lambda \sum_{\pm \mp }\sum_{n^{\prime }m^{\prime
},j^{\prime }k^{\prime }}\alpha \frac{g_{j}\left( \sqrt{\left( j+1\right) j}+%
\sqrt{\left( j-1\right) j}\right) -g_{k}\left( \sqrt{\left( k+1\right) k}+%
\sqrt{\left( k-1\right) k}\right) }{\left( \varepsilon _{n^{\prime },k}^{\pm
\pm }-\varepsilon _{m^{\prime },j}^{\mp \mp }\right) -\left( \varepsilon
_{n,k}^{\pm \pm }-\varepsilon _{m,j}^{\mp \mp }\right) }}{\left\{ \left(
\varepsilon _{n,k}^{\pm \pm }-\varepsilon _{m,j}^{\mp \mp }\right) -\lambda
\sum_{\pm \mp }\sum_{n^{\prime }m^{\prime },j^{\prime }k^{\prime }}\frac{%
g_{j}\left( \sqrt{\left( j+1\right) j}+\sqrt{\left( j-1\right) j}\right)
-g_{k}\left( \sqrt{\left( k+1\right) k}+\sqrt{\left( k-1\right) k}\right) }{%
\left( \varepsilon _{n^{\prime },k}^{\pm \pm }-\varepsilon _{m^{\prime
},j}^{\mp \mp }\right) -\left( \varepsilon _{n,k}^{\pm \pm }-\varepsilon
_{m,j}^{\mp \mp }\right) }\right\} }.  \nonumber
\end{eqnarray}%
This shows that: $\left( 1\right) $ the change of the eigenvalues after
loading the coupling interaction, introduces a phase error in the swap
operator, $\left( 2\right) $ the error can easily be cancelled by adjusting
the coupling time ($\triangle t_{sw}$) based on $t_{sw}$, since the
eigenvectors are invariant. This demonstrates an important advantage of
performing quantum Logical operations in the projected subspace.

Moreover, from Eqs.(\ref{eqn9}), (\ref{eqn10}) and (\ref{eqn60}) it is
evident that eigenvalue $E_{nm,kj}^{\pm \pm \mp \mp }$ may be complex (in
fact, this model is a type of Friedrichs model, possessing a complex
spectrum in the sense of extending Hilbert space$^{\left[ 22-24\right] }$).
This means that the time evolution of $\rho _{nm,kj}^{\pm \pm \mp \mp }$ in
Eq.(\ref{eqn55aa}) may be asymmetric. Thus, a Rigged Liouville Space (RLS)
formulation can be chosen$^{\left[ 5,6\right] }$ to describe the subsystem,
since a Hilbert space formulation provides an inadequate description. The
evolution of the projected density operator and its adjoint operator are
expressed as 
\begin{eqnarray}
\sum_{\pm \mp }\sum_{nmjk}\left| \rho _{nm,kj}^{\pm \pm \mp \mp }\left(
t\right) \right) &=&\sum_{\pm \mp }\sum_{nmjk}e^{-iE_{nm,kj}^{\pm \pm \mp
\mp }t}\left( \phi _{nm,jk}^{\pm \pm \mp \mp }\right. \left| \rho
_{nm,kj}^{\pm \pm \mp \mp }\left( 0\right) \right) \left| \phi _{nm,jk}^{\pm
\pm \mp \mp }\right) ,  \label{eqn69} \\
\sum_{\pm \mp }\sum_{nmjk}\left( \widetilde{\rho }_{nm,kj}^{\pm \pm \mp \mp
}\left( t\right) \right| &=&\sum_{\pm \mp }\sum_{nmjk}e^{iE_{nm,kj}^{\pm \pm
\mp \mp }t}\left( \widetilde{\rho }_{nm,kj}^{\pm \pm \mp \mp }\left(
0\right) \right| \left. \phi _{nm,jk}^{\pm \pm \mp \mp }\right) \left( \phi
_{nm,jk}^{\pm \pm \mp \mp }\right| ,  \nonumber
\end{eqnarray}%
where $\rho _{nm,kj}^{\pm \pm \mp \mp }\left( t\right) $ may exist in the
test space $\Phi _{SR}\otimes \Phi _{SR}$, which is a dense subspace of the
Liouville Space, representing the physical states which can be prepared in
an actual experiment. Its adjoint $\widetilde{\rho }_{nm,kj}^{\pm \pm \mp
\mp }\left( t\right) $ lies in the dual space $^{\times }\Phi _{SR}\otimes
\Phi _{SR}^{\times }$, representing a procedure that associates with each
state a number, while preserving the linear structure which results from the
superposition principle. This is a RLS structure which facilitates
describing irreversible processes like decoherence and dissipation due to
interaction with the environment. This poses an interesting question as to
what is new in this formulation for an open quantum system,\ if it performs
quantum logical operations in a RLS (RHS). For convenience we assume that $%
\left| \rho _{nm,kj}^{\pm \pm \mp \mp }\left( t\right) \right) $, $\left( 
\widetilde{\rho }_{nm,kj}^{\pm \pm \mp \mp }\left( t\right) \right| $ are
biorthonormalized with respect to each other, and the evolution of $\rho
_{nm,kj}^{\pm \pm \mp \mp }\left( t\right) $ (in Eq.(\ref{eqn55aa})) in the
projected subspace permits the system to perform quantum computing in a RLS
(RHS).

Indeed, if we construct a general quantum logical operator ${\cal Q}$ given
by some function $f$ of the biorthonormal states $\left\{ \left| \rho
_{nm,kj}^{\pm \mp }\left( t\right) \right) \text{, }\left( \widetilde{\rho }%
_{nm,kj}^{\pm \mp }\left( t\right) \right| \right\} $ in a RLS as 
\begin{eqnarray}
{\cal Q}g\left( \left| \rho _{nm,kj}^{\pm \mp }\left( t\right) \right)
\right) &=&g^{\prime }\left( \left| \rho _{nm,kj}^{\prime \pm \mp }\left(
t\right) \right) \right) ,  \label{eqn70} \\
h\left( \left( \widetilde{\rho }_{nm,kj}^{\pm \mp }\left( t\right) \right|
\right) {\cal Q} &=&h^{\prime }\left( \left( \widetilde{\rho }%
_{nm,kj}^{\prime \pm \mp }\left( t\right) \right| \right) ,  \label{eqn71}
\end{eqnarray}%
and 
\begin{equation}
{\cal Q\equiv }f\left( \left| \rho _{nm,kj}^{\pm \mp }\left( t\right)
\right) \left( \widetilde{\rho }_{nm,kj}^{\pm \mp }\left( t\right) \right|
\right) .  \label{eqn72}
\end{equation}%
Then the states or observables are $g\left( \left| \rho _{nm,kj}^{\pm \mp
}\left( t\right) \right) \right) $, $g^{\prime }\left( \left| \rho
_{nm,kj}^{\prime \pm \mp }\left( t\right) \right) \right) \in \Phi
_{S}\otimes \Phi _{S}$, $h\left( \left( \widetilde{\rho }_{nm,kj}^{\pm \mp
}\left( t\right) \right| \right) $, $h^{\prime }\left( \left( \widetilde{%
\rho }_{nm,kj}^{\prime \pm \mp }\left( t\right) \right| \right) \in $ $%
^{\times }\Phi _{S}\otimes \Phi _{S}^{\times }$, and the space is closed to
the dual pair spaces $\Phi _{S}\otimes \Phi _{S}$ and $^{\times }\Phi
_{S}\otimes \Phi _{S}^{\times }$. The RLS structure is $\Phi _{S}\otimes
\Phi _{S}\subset {\cal H}\otimes {\cal H}\subset $ $^{\times }\Phi
_{S}\otimes \Phi _{S}^{\times }$. For example, a quantum universal gate,
such as a two qubit based quantum Controlled-Not logical gate (XOR gate) in
RLS, can generally be constructed from a combination of the projection
operators as: 
\begin{equation}
{\cal CN}=\left| 00\right) \widetilde{\left( 00\right| }+\left| 01\right) 
\widetilde{\left( 01\right| }+\left| 10\right) \widetilde{\left( 11\right| }%
+\left| 11\right) \widetilde{\left( 10\right| },  \label{eqn73}
\end{equation}%
which induces the corresponding quantum logical operations, 
\begin{eqnarray}
{\cal CN}\left| 00\right) &=&\left| 00\right) ,{\cal CN}\left| 01\right)
=\left| 01\right) ,  \label{eqn74} \\
{\cal CN}\left| 10\right) &=&\left| 11\right) ,{\cal CN}\left| 11\right)
=\left| 10\right) ,  \nonumber \\
\widetilde{\left( 00\right| }{\cal CN} &=&\widetilde{\left( 00\right| },%
\widetilde{\left( 01\right| }{\cal CN}=\widetilde{\left( 01\right| }, 
\nonumber \\
\widetilde{\left( 10\right| }{\cal CN} &=&\widetilde{\left( 11\right| },%
\widetilde{\left( 11\right| }{\cal CN}=\widetilde{\left( 10\right| }, 
\nonumber
\end{eqnarray}%
and demonstrates that the quantum logical operations are closed in RHS. The
property of closure for quantum logical operators in RLS (RHS) is important
for practical applications. This shows that quantum logical operations in
RLS (RHS) inherit the essential properties of those in Liouville (Hilbert)
space, if the biorthornomal property of generalized eigenstates still hold.

One may still argue that the advantages of the above approach for quantum
logical operations in RLS (RHS), for an open computing system $S$, are not
evident. We would argue that the Liouville (Hilbert) space approach
generally fails to describe open quantum computing systems which undergo
irreversible processes. These include semigroup evolution and
non-self-adjoint Hamiltonians of subsystems, far from equilibrium. This is
something that the RLS (RHS) formulation is well suited to handling. To
construct quantum logical operations in projected subspaces, the influence
of decoherence induced by changes in the eigenvalues, may easily be
cancelled since the eigenvectors remain invariant. Moreover, in this space
the evolution of the states are permitted to be time asymmetric, providing a
framework for describing the irreversibility of practical open systems. This
irreversibility does not change quantum reversible logical operations to
quantum irreversible logical operation in the quantum universal
Controlled-Not logical gate. To appreciate this, one must distinguish
between irreversibility of a quantum logical operation, introduced by the
structure of logical gate and irreversibility of the process, induced by
interactions with the environment. Reversible computation means reversible
logical operations on the structure of the logical gate. In this sense,
quantum computing in RLS (RHS) does not adversely impact on reversible
quantum logical operations and permits computing any reversible function,
although irreversible processes exist. This is further clarified below in
relation to construction of a quantum Turing machine in RHS.

\section{Construction of a Turing Machine in RHS}

Consider, in general, an open system with $N$ components (atoms) each of
which can have two states. We design a quantum Turing machine consisting of $%
N$ $=n+1$ pseudospins or levels $\left| \pm \left( j\right) \right\rangle $, 
$j=h,1,2,\cdots ,n,$ in a $2^{n+1}$ dimensional test space $\Phi $ and $%
\left\langle \widetilde{\pm \left( j\right) }\right| ,$ $j=h,1,2,\cdots ,n,$
in the dual space of the test space $\Phi ^{\times }$ with biorthnomality
and isometry, 
\begin{equation}
\left\langle \widetilde{\pm \left( j\right) }\right| \left. \mp \left(
j^{\prime }\right) \right\rangle =\delta _{\pm \mp }\delta _{jj^{\prime }},
\label{eqn75}
\end{equation}%
\begin{equation}
\left\langle U^{+}\widetilde{\pm \left( j\right) }\right| \left. U\left( \pm
\left( j\right) \right) \right\rangle =\left\langle \widetilde{\pm \left(
j\right) }\right| \left. \pm \left( j\right) \right\rangle ,  \label{eqn76}
\end{equation}%
where the Turing head is denoted as $h$, and the Turing tape pseudospins are
denoted as $1,\cdots ,n$, and $\pm $ represent two states or two
observables. The quantum Turing machine can be considered as an open system
in which the Turing tape is a special finite bath coupled to the Turing
head. The state of the quantum Turing machine $\left| \psi \right\rangle $
lies in a $2^{n+1}$ dimensional dense subspace $\Phi $ of the Hilbert Space
and is spanned by $\left\{ \left| \pm \left( j\right) \right\rangle \right\} 
$, and its observables $\left\langle \widetilde{\psi }\right| $ are spanned
by $\left\{ \left\langle \widetilde{\pm \left( j\right) }\right| \right\} $
in the dual space $\Phi ^{\times }$. The evolution of a pure state $\left|
\psi _{0}\right\rangle \left\langle \widetilde{\psi }_{0}\right| $ is given
by 
\begin{equation}
\left| \psi _{j}\right\rangle \left\langle \widetilde{\psi }_{j}\right|
=U\left( j\right) \left| \psi _{0}\right\rangle \left\langle \widetilde{\psi 
}_{0}\right| \widetilde{U}\left( j\right) ^{+},  \label{eqn77}
\end{equation}%
where the non-unitary evolution operator $U$ can be expanded by the
non-self-adjoint generators defined by

\begin{eqnarray}
\widehat{\lambda }_{x}\left( j\right) &=&P_{01}\left( j\right) +P_{10}\left(
j\right) ,  \label{eqn78} \\
\widehat{\lambda }_{y}\left( j\right) &=&iP_{01}\left( j\right)
-iP_{10}\left( j\right) ,  \nonumber \\
\widehat{\lambda }_{z}\left( j\right) &=&P_{11}\left( j\right) -P_{00}\left(
j\right) ,  \nonumber
\end{eqnarray}%
with 
\begin{equation}
P_{ik}\left( j\right) =\left| i\left( j\right) \right\rangle \left\langle 
\widetilde{k}\left( j\right) \right| .  \label{eqn79}
\end{equation}

Following ref.$\left[ 25\right] $, we restrict ourselves to the Bloch vector 
$\overrightarrow{{\bf \lambda }}$ of the Turing head $h$ as 
\begin{equation}
\lambda _{i}^{h}=\left\langle \widetilde{\psi }_{h}\right| \widehat{\lambda }%
_{i}\left( h\right) \otimes 1\left( 1\right) \otimes \cdots \otimes 1\left(
n\right) \left| \psi _{h}\right\rangle .  \label{eqn80}
\end{equation}%
Hence, for $2^{n}$ biorthonormal initial tape states $\left| \phi
_{0}\right\rangle =\left| \pm \left( 1\right) \right\rangle \otimes \left|
\pm \left( 2\right) \right\rangle \otimes \cdots \otimes \left| \pm \left(
n\right) \right\rangle ,$ $\left\langle \widetilde{\phi }_{0}\right|
=\left\langle \widetilde{\pm \left( n\right) }\right| $ $\otimes \cdots
\otimes \left\langle \widetilde{\pm \left( 2\right) }\right| \otimes
\left\langle \widetilde{\pm \left( 1\right) }\right| $ and arbitrary Turing
head $h$ states $\left| \varphi _{h}\right\rangle ,$ $\left\langle 
\widetilde{\varphi }_{h}\right| $, the states of Turing machine $\left| \psi
_{h}\right\rangle $, $\left\langle \widetilde{\psi }_{h}\right| $ can be
written as the tensor product states 
\begin{eqnarray}
\left| \psi _{h}\right\rangle &=&\left| \varphi _{h}\otimes \phi
_{0}\right\rangle ,  \label{eqn81} \\
\left\langle \widetilde{\psi }_{h}\right| &=&\left\langle \widetilde{\varphi 
}_{h}\otimes \widetilde{\phi }_{0}\right| ,  \nonumber
\end{eqnarray}%
and hence the Bloch vector of the Turing head (\ref{eqn80}) is described by
the above tensor product states as 
\begin{equation}
\lambda _{i}^{h}\left( \phi _{0},\widetilde{\phi }_{0}\right) =\left\langle 
\widetilde{\varphi }_{h}\otimes \widetilde{\phi }_{0}\right| \widehat{%
\lambda }_{i}\left( h\right) \otimes 1\left( 1\right) \otimes \cdots \otimes
1\left( n\right) \left| \varphi _{h}\otimes \phi _{0}\right\rangle
\label{eqn82}
\end{equation}%
performing a pure state trajectory on the Bloch circle, 
\begin{equation}
\lambda _{y}^{h}\left( \phi _{0},\widetilde{\phi }_{0}\right) ^{2}+\lambda
_{z}^{h}\left( \phi _{0},\widetilde{\phi }_{0}\right) ^{2}=1.  \label{eqn83}
\end{equation}%
Any initial state of the Turing machine $\left| \psi _{0}\right\rangle $, $%
\left\langle \widetilde{\psi }_{0}\right| $ with the Turing head $h$ in the
pure state $\left| \varphi _{0}\left( h\right) \right\rangle $, $%
\left\langle \widetilde{\varphi }_{0}\left( h\right) \right| $ can thus be
written 
\begin{equation}
\left| \psi _{0}\right\rangle =\sum_{j=1}^{2^{n}}a_{j}\left| \varphi
_{0}\left( h\right) \otimes \phi _{0}^{j}\right\rangle ,  \label{eqn84}
\end{equation}%
\[
\left\langle \widetilde{\psi }_{0}\right|
=\sum_{j=1}^{2^{n}}b_{j}\left\langle \widetilde{\varphi }_{0}\left( h\right)
\otimes \widetilde{\phi }_{0}^{j}\right| . 
\]%
This yields a motion of the Bloch vector of Turing head $h$ induced by the
change of the tape states, given by

\begin{equation}
\lambda _{k}^{h}\left( \psi _{0},\widetilde{\psi }_{0}\right)
=\sum_{j=1}^{2^{n}}a_{j}b_{j}\lambda _{k}^{h}\left( \phi _{0}^{j},\widetilde{%
\phi }_{0}^{j}\right) ,  \label{eqn85}
\end{equation}%
where $a_{j}$, $b_{j}$ are expansion coefficients. This shows that the
motion of the Bloch vector of Turing head $\lambda _{k}^{h}\left( \psi _{0},%
\widetilde{\psi }_{0}\right) $ constructed in RHS can be expressed as $%
2^{n}-1$ entangled Bloch vectors of the Turing head $\lambda _{k}^{h}\left(
\phi _{0}^{j},\widetilde{\phi }_{0}^{j}\right) $ induced by a pure Turing
head state $\left| \varphi _{0}\left( h\right) \right\rangle $, $%
\left\langle \widetilde{\varphi }_{0}\left( h\right) \right| $, and $2^{n}$
initial tape states $\left| \phi _{0}\right\rangle $, $\left\langle 
\widetilde{\phi }_{0}\right| $. This means that the motion of the Turing
head $h$, $\lambda _{k}^{h}\left( \psi _{0},\widetilde{\psi }_{0}\right) $,
can perform many parallel primitive trajectories of the Bloch vector of the
Turing head $h$, $\lambda _{k}^{h}\left( \phi _{0}^{j},\widetilde{\phi }%
_{0}^{j}\right) $, identical to the situation of the reversible Turing head
constructed in the Hilbert Space formulation$^{\left[ 25\right] }$. This is
because the generalized basis in RHS remains biorthonormal and isometric,
corresponding to the orthonormal property, and invariance of the inner
product for the basis in Hilbert Space. This enables most properties of the
quantum Turing machine in Hilbert Space to be inherited in RHS, with the
exception of the reversible property of an ideal process.

\section{Conclusions}

In conclusion, a subdynamic formulation for an open quantum system is
presented using a simpler approach than discussed previously. Based on the
subdynamical kinetic equation for an open quantum system, the properties of
coherence and decoherence-free conditions for three types of interactions
were analyzed. A condition for a quantum computing system to remain in
coherent states in the projected subspaces is found. That is, the
interactions between the system and its environment should be ``diagonal''
or ``triangular''. Moreover, one can find universal DF projected subspaces
by using the subdynamical kinetic equation without restrictions on the type
of decoherence and without introducing any approximations, although the
total space may be decoherent. An implied universal concept in our analysis
is that the invariant eigenvectors remain a DF structure for any stationary
states of the system and that changes of the eigenvalues only induce a phase
shift. This type of phase shift which may introduce a type of phase error
for the evolved states does not change the DF property of the subspace. The
possible phase error can be cancelled, due to the invariance of the
eigenvectors in the projected subspace under the assumption of a homogeneous
distribution of the energy shifts. In this projected subspace, one can
construct a quantum XOR operator and show that the evolution of the SWAP
operator is time asymmetric. Hence, we propose a formulation for performing
quantum computing in RLS (RHS) by constructing a general quantum
Controlled-Not logical gate, with corresponding operations in RLS (RHS), and
a generalized quantum Turing machine in RHS.

{\LARGE \ }

{\LARGE \ }

{\LARGE \ }

{\LARGE \ }

{\LARGE \ }

{\LARGE \ }

\bigskip {\bf ACKNOWLEDGEMENTS}

{\LARGE \ }

We gratefully acknowledge financial support from NSERC, MITACS, CIPI, MMO,
CITO and China State Key Projects of Basic Research and Natural Science
foundation (G1999064509, N$_{0}$ 79970121, 60072032).

\end{document}